\documentclass[final,5p,times,twocolumn]{elsarticle}

\usepackage{amssymb}
\usepackage{graphicx}
\usepackage{subcaption}
\usepackage{amsmath}
\usepackage{nccmath}
\usepackage{amsfonts}
\DeclareGraphicsExtensions{.png}
\usepackage{multirow}
\usepackage[table]{xcolor}
\usepackage{textcomp}
\usepackage{url}

\journal{JNM}

\begin{document}

\begin{frontmatter}


\title{Decoding Imagined Speech and Computer Control using Brain Waves}
\author{Abhiram Singh\corref{cor1}}
\ead{abhiram25.1990@gmail.com}
\author{Ashwin Gumaste\corref{cor2}}
\ead{ashwin@ashwin.name}
\cortext[cor1]{Corresponding author}
\address{Department of Computer Science and Engineering, Indian Institute of Technology Bombay, India}





\begin{abstract}
\textit{Background}. In this work, we explore the possibility of decoding Imagined Speech (IS) brain waves using machine learning techniques. \textit{Approach}. We design two finite state machines to create an interface for controlling a computer system using an IS-based brain-computer interface. To decode IS signals, we propose a covariance matrix of Electroencephalogram channels as input features, covariance matrices projection to tangent space for obtaining vectors from matrices, principal component analysis for dimension reduction of vectors, an artificial neural network (ANN) as a classification model, and bootstrap aggregation for creating an ensemble of ANN models. \textit{Result}. Based on these findings, we are first to use an IS-based system to operate a computer and obtain an information transfer rate of 21-bits-per-minute. The proposed approach can decode the IS signal with a mean classification accuracy of 85\% on classifying one long vs. short word. Our proposed approach can also differentiate between IS and rest state brain signals with a mean classification accuracy of 94\%. \textit{Comparison}. After comparison, we show that our approach performs equivalent to the state-of-the-art approach (SOTA) on decoding long vs. short word classification task. We also show that the proposed method outperforms SOTA significantly on decoding three short words and vowels with an average margin of 11\% and 9\%, respectively. \textit{Conclusion}. These results show that the proposed approach can decode a wide variety of IS signals and is practically applicable in a real-time environment.
\end{abstract}




\end{frontmatter}

\section{Introduction}
Brain-Computer Interface (BCI) is a combination of hardware (used to capture brain signals) and software (to analyze and understand different cognitive tasks). Research in BCI is getting popular to study human behavior, diagnose brain diseases, and utilize as a human-computer interface (HCI) device. A BCI system can be seen as a replacement for existing technologies such as a touch screen, mouse, or keyboard. Many BCI systems utilize different paradigms such as P300 or motor imaginary for Human-Computer Interaction \cite{wolpaw}, \cite{Mugler2010}.

\textit{Imagined Speech}: Various activities generate electrical signals from the brain. Imagined speech (IS) or speech imaginary \cite{Hickok2007TheCO} is one such class of brain signals in which the user speaks in the mind without explicitly moving any articulators. IS is different from silent speech, in which a user thinks to move articulators during the imagination of words. Hence, silent speech is likely to generate signals from the brain's motor cortex, whereas IS generates the signals from Broca's and Wernicke's areas \cite{Sahin2009SequentialPO}, \cite{Martin2016}.

\textit{Electroencephalography}: There exist different techniques to capture electrical signals from the brain. Electroencephalography (EEG) \cite{Michal2002EEG} is one such widely used technique that involves placing electrodes over the scalp in a non-invasive fashion. These electrodes capture voltage differences generated due to ion movement along the brain neurons. These measurements are obtained over a time period to form an EEG signal. The number of electrodes can vary from sparse (just 1) to dense (256), determined based on the application requirements. The EEG signal requires preprocessing steps (e.g., band-pass filtering, artifacts removal) before extracting the useful information. After the noise removal, relevant features (temporal, spectral, and spatial) are extracted from the EEG signal and provided as an input to a classifier. The classifier categorizes input features into one of the classes, where each class represents an imagined task.

\textit{EEG signal decoding}: To automate the preprocessing and features extraction steps with the EEG signal classification, Lawhern et al. \cite{eegnet} proposed a deep convolutional neural network (\textit{EEGNet}). The performance of \textit{EEGNet} was evaluated on four datasets, where each dataset represents a different BCI classification task. To extract both spectral and temporal features of the EEG signal, Wang et al. \cite{wang2017simulation} proposed CNN and RNN based deep learning model for decoding the imagined speech signals on the synthetic EEG data. Similarly, Bashivan et al. \cite{bashivan2015learning} proposed a recurrent-convolution-based deep learning model for preserving temporal, spatial, and spectral information present in the EEG signal. In a similar work, Zhang et al. \cite{zhang2018cascade} proposed a combination of convolution and recurrent model (interconnected in a cascade and parallel fashion) for decoding the EEG signals. In the context of machine learning techniques, Nguyen et al. \cite{Nguyen2018IS} proposed features from tangent space with relevance vector machine. Tomioka et al. \cite{Blankertz2008} utilized a common spatial pattern to extract log-variance features of the EEG signal with the linear discriminant analysis as a classification model. In a similar work, Dasalla et al. \cite{Dasalla2009} applied a common spatial pattern with the support vector machine to classify the EEG signals. Min et al. \cite{Min2016} utilized statistical features as an input to the extreme learning machine for decode the EEG signals.

\textit{Proposed work}: This paper focuses on the design, implementation of an IS-based BCI system, and classification of EEG-based IS signals. The reason behind using IS signals in BCI is because the speech-based system is much faster, expected to take less training time, provides more comfort than motor imaginary tasks, and leads to a natural way of HCI \cite{Sherry2016}. Therefore, IS signals may lead to an overall improved user experience in computer interaction. The work in this paper assumes that the data is not fully corrupted with noise. Subjects participating in IS experiments are instructed to follow specific guidelines, making this assumption feasible (though this may not always be true in a real-life scenario). So, there is a possibility of extracting useful information related to the imagined speech and, after that, decode the signal.
The proposed work aims to identify the discriminative features and a classification model that improves decoding performance on different IS tasks and is robust to noise. Based on experimental results, we propose Tangent Space (TS) \cite{Barachant2012Riemann} as input features to an Artificial Feed Forward Neural Networks (ANN) \cite{Goodfellow2016DL} model. 

\textit{Results}: 
We tested one IS-based computer control design in a partial online setting and obtained an information transfer rate (ITR) of 21 bits/minute.
For decoding the IS signal, our proposed approach improves the mean classification accuracy from 49.3\% to 60.35\%, 49.2 \% to 58.61\%, 66.56\% to 69.43\%, and 73.27\% to 78.51\% on three short words, three vowels, two long words, and one long vs. one short word classification tasks respectively, from the state of the art.

\textit{Problem Statement}: Given an EEG signal, we desire to identify whether it belongs to an imagined speech category. If so, then we desire to decode the actual, imagined word or word category. Subsequently, we want to use the decoded information to take appropriate action for computer interaction.

\textit{Contribution}: Our work leads to the following contributions:
\begin{enumerate}
\item We propose an FSM to operate the computer system by using only IS signals. Thereafter, we develop a new and simple graphical user interface (GUI) corresponding to the FSM for user interaction with the system. To the best of our knowledge, this is the first-ever approach for general-purpose computer control, which is based only on the IS signals. We discuss two FSM designs for binary classification tasks using IS signals and then focus on several improvements to build a fully functional system that can work in a real-time (online) setting. For demonstration, we simulate an FSM design on a publicly available dataset and obtain an ITR of 21-bits-per minute. 

\item We consider the aspect of generalization of neural networks (NN) on IS signals. We identify i) The covariance matrix as the most useful discriminative feature; ii) Tangent Space (TS) as discriminative information preservative transformation of the covariance matrix to vectors; iii) PCA as a dimension reduction technique; iv) Artificial Neural Network (NN) as the most successful classification model with boot-strap aggregation (bagging) for combining the output of multiple NN. Results confirm that our proposed approach improves the classifier performance significantly over the existing machine learning and two baseline deep learning models (convolutional neural network (CNN) and recurrent neural network (RNN).

\item We show that the proposed approach can discriminate IS-based EEG signals from participant's rest state EEG signals. This step helps to eliminate non-IS signals.
\end{enumerate}

\textit{Paper organization}: This paper is organized as follows. Section 2 describes the user interface designs and pipeline for an IS based real-time BCI system for computer control. Section 3 provides the dataset details and describes the proposed approach for feature extraction and classification. Section 4 shows the results of the proposed approach and comparison with deep learning and machine learning models. Section 5 provides discussion, related work and conclusion.

\section{Imagined speech for computer interaction}
In this section, we provide design and implementation details of the imagined speech-based computer interaction system. We propose two designs: 1) creating a new Graphics User Interface (GUI) to click anywhere on a computer screen; 2) a design that utilizes Arrow, Enter, and Backspace keys of a keyboard to perform corresponding actions on a computer.

\begin{figure*}[t]
    \centering
    \begin{subfigure}[t]{0.5\textwidth}
        \includegraphics[width=0.85\textwidth,height=5cm]{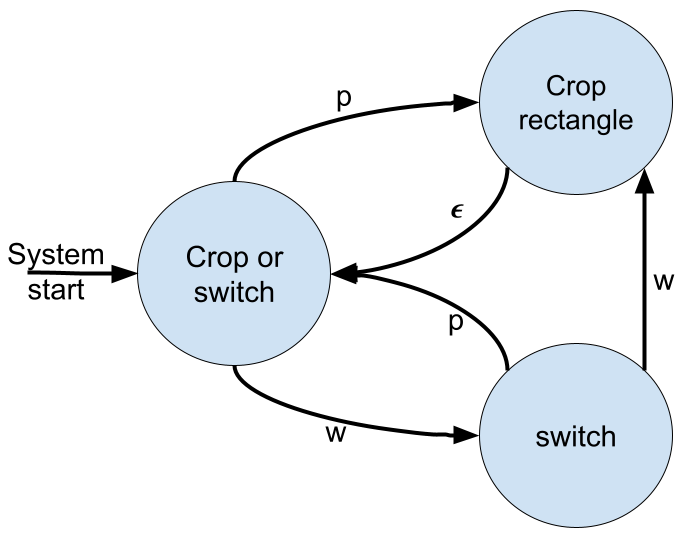}
        \centering
        \caption{}
        \label{stateDiagram1}
    \end{subfigure}%
    \begin{subfigure}[t]{0.5\textwidth}
        \includegraphics[width=0.95\textwidth,height=5cm]{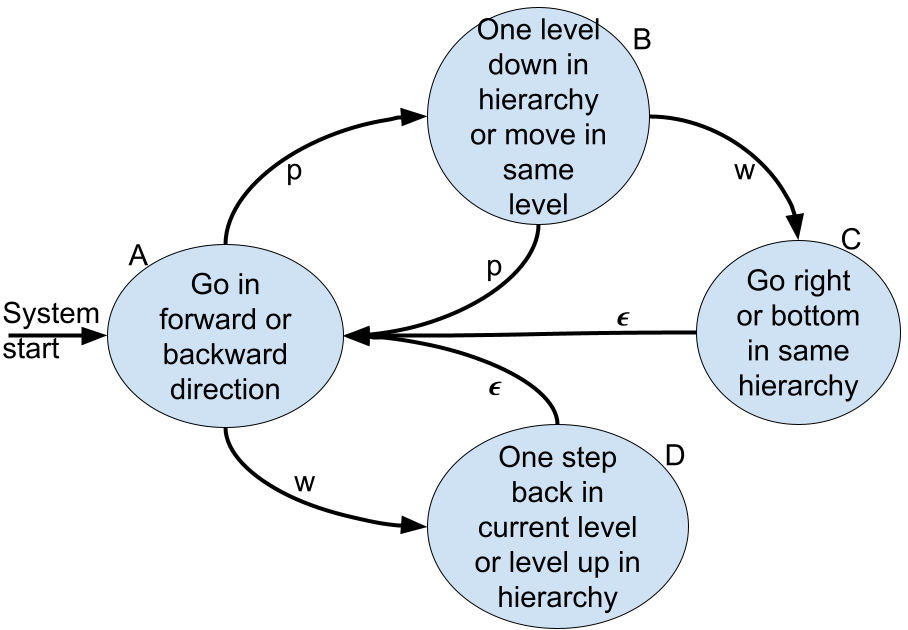}
        \centering
        \caption{}
        \label{stateDiagram2}
    \end{subfigure}
    \caption{(a) State diagram 1 of computer control application. This FSM design can be used to click anywhere on a computer screen. In the \textit{crop or switch} state,  a user has an option to select one of the \textit{crop rectangle} or \textit{switch} states by thinking of a short or long word. In the \textit{crop rectangle} state, the user narrows down the visible rectangular window and in the \textit{switch} state, the user can switch the rectangular window either for double click operation (thinking a short word) on a folder or recreate a rectangle from the previous crop operation (thinking a long word). (b) State diagram 2 of computer control application. This FSM design is used to navigate within a file system. In the figure, p, w and $\epsilon$ stands for a short word, long word, and transition without user input.}
    \label{stateDiagram}
\end{figure*}

\subsection{Design}
To control a computer, the first requirement is to locate desired content displayed on a computer screen. So there must be some provision with which a user can reach the target location. Currently, this step is the movement of a mouse shown as a change in the cursor position on the screen. A keyboard may also be used by using the Tab or arrow keys to reach the target. Since cursor control requires continuous input from the user, and the imagined speech classifier provides a discrete output; therefore it does not make sense to control continuous movement using discrete steps. Hence, the type of classifier output (continuous or discrete) must be considered in the design.

\subsubsection{Assumption} The binary classifier is used for imagined speech decoding. This assumption is due to simplicity in GUI demonstration and high classification accuracy obtained by the binary classifier (\S \ref{sec:results}) on the dataset (\ref{sec:data_approach}). Also, the classifier is trained such that the 0 output corresponds to a short word, and the 1 output corresponds to a long word. These assumptions can be relaxed by increasing the classifier performance on the multiclass classification problem. 

\subsubsection{First design} 
In the first design of FSM (Figure \ref{stateDiagram1}), we propose the following steps in each iteration to open a folder currently being displayed on the screen.

\begin{figure*}[t]
    \centering
    \begin{subfigure}[t]{0.5\textwidth}
        \includegraphics[width=0.9\textwidth,height=4cm]{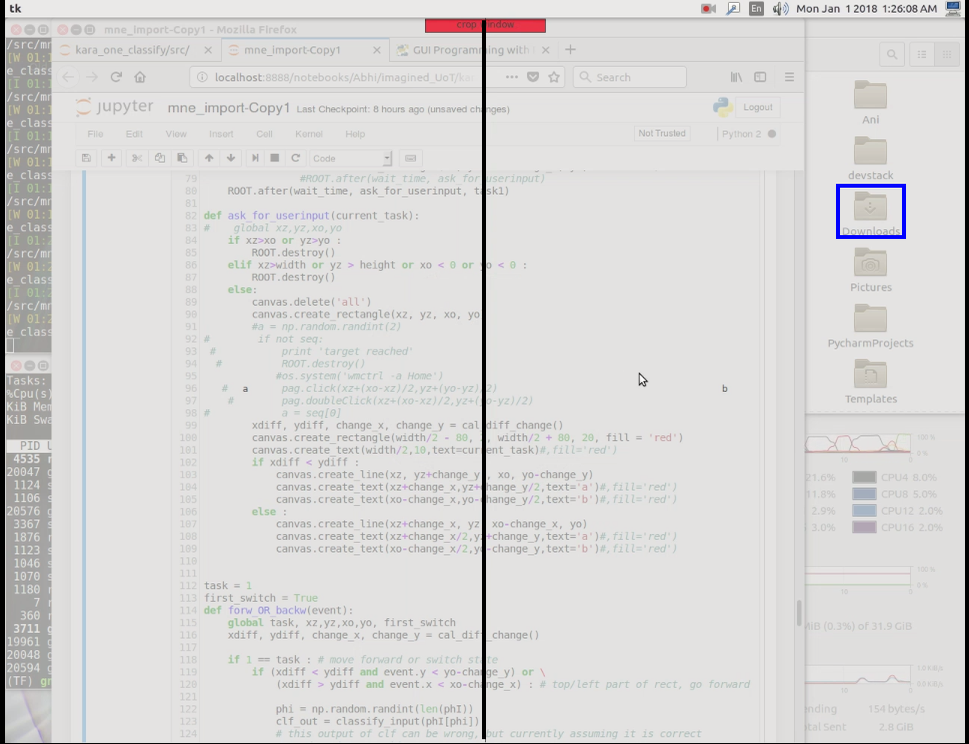}
        \centering
        \caption{}
        \label{crop1}
    \end{subfigure}%
    \begin{subfigure}[t]{0.5\textwidth}
        \includegraphics[width=0.9\textwidth,height=4cm]{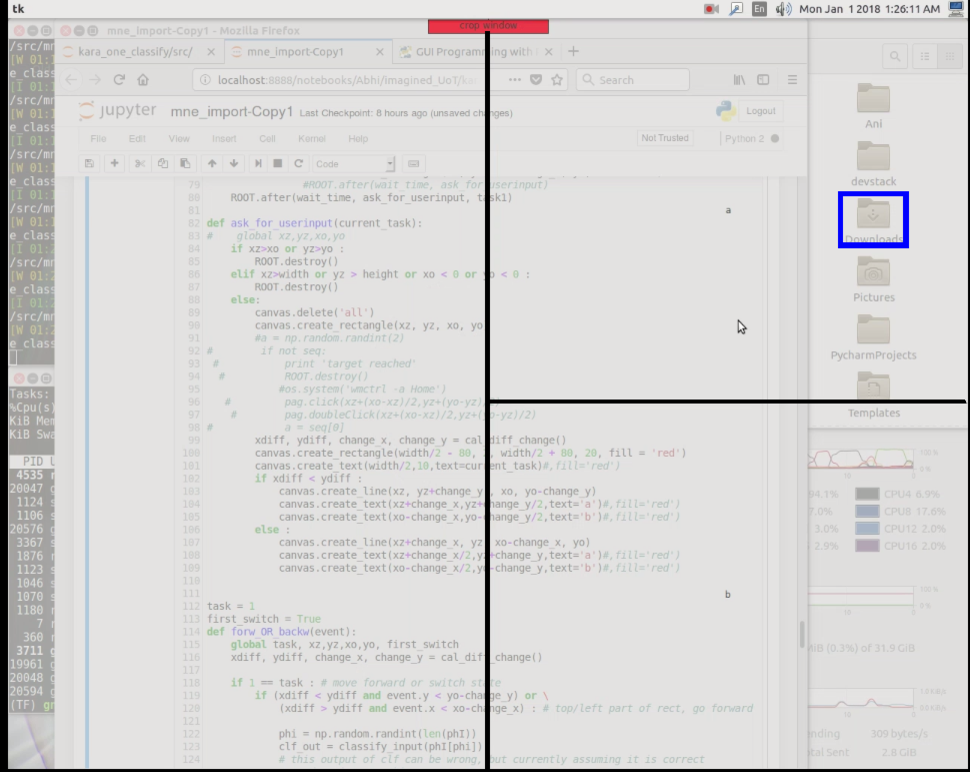}
        \centering
        \caption{}
        \label{crop2}
    \end{subfigure}
    \caption{(a) The first division of rectangle from crop action. To open a folder in the blue box, the user thinks of one long word so that the right part of the rectangle is selected and the left part is cropped. (b) The second division of rectangle from crop action. To open a folder in the blue box, the user thinks of one short word so that the top part of the rectangle is selected and the bottom part is cropped.}
    \label{crop}
\end{figure*}

\textbf{1)} We obtain the screen resolution and create a rectangular window with partial transparency of the same size as that of the screen resolution. We then divide the current rectangle into two halves (as shown in Figure \ref{crop1}). If the length is greater or equal to the breadth, then we divide the rectangle along the length, otherwise divide the rectangle along the breadth. 

\textbf{3)} We then display one short word on one half of the rectangle and one long word on the other side of the rectangle. For consistency, if the rectangle is divided along its length, then the short word is always displayed on the left part of the rectangle, and the long word is always displayed on the right part of the rectangle. Similarly, if the rectangle is divided along its breath, then the short word is displayed on the top part, and the long word is displayed on the lower part of the rectangle. The short and long words are chosen randomly from their respective sets. 

\textbf{4)} A display response is used for ensuring that the user starts thinking of either the short or long word in a given time-interval leading to the capture of the corresponding brain signals. A response is provided by showing one textbox for each rectangle and a textbox specifies a word to be imagined. For example, if the rectangle is divided vertically, then a short word is displayed on the left part, and a long word is displayed on the right part of the rectangle. The user imagines either a short or long word by looking at the part of the rectangle under which the target folder is located. 

\textbf{5)} After capturing the IS signal, it is pre-processed, features are extracted and given to the classifier to decode the word imagined by the user. If a classifier generates an output of 0 then, the rectangle part (either right or bottom) representing a long word is removed. In contrast, if the output is a 1 then, the rectangle part (either left or top) corresponding to a short word is removed. 

\textbf{6)} Steps 2-5 are repeated until the rectangle becomes small enough to cover the folder entirely. At this stage, the user needs to switch the window and double-click on the folder. However, until this stage, the system only recognizes one action: to crop the current window to reach a target location. To introduce another action in the design, at the starting of each step, two options are displayed to the user. The first option asks a user to go to the crop state, whereas the second option ask for a switch state. A user selects the crop action by thinking of a short word, and this leads to the system state where all the above-defined steps 1-5 can be performed. 

\textbf{7)} A user can select the switch option by thinking of a long word. If at a switch state, the user at any time thinks of a short word then the system switches the window and double clicks the folder behind the current rectangle. Thereafter, the system resets its state, the rectangle is set to the full-screen resolution, and the whole process restarts to select a different folder. However, if the classifier made a mistake on the previous crop, the user can go to a switch state, think of a long word to recreate the last crop's rectangle, and go to the crop rectangle state again.

\subsubsection{Second design} 
The second design (shown by an FSM in Figure \ref{stateDiagram2}) converts user imagined speech into the keyboard actions. Here we demonstrate one application to utilize a tree file directory structure. The tree structure can be divided into multiple levels, with a root at the top and leaves at the bottom. A file in the system represents a tree leaf, and the root is the top level directory of the computer system. Initially, a user decides to open a particular file in the computer system. Then computer control is shifted to the root of a directory tree. There might be multiple directories at the root. So the first among them is selected. Based on the target file location, the user can either navigate at the same level of the tree or go a level down. To achieve this, a user imagines of one short word to change the system state from A to B (see Figure \ref{stateDiagram2}). In-state B, the user can either move in the same directory level or go one level down. A user can think of a short word to go one level down along the directory tree hierarchy. This user's imagined speech is converted into action corresponding to pressing an Enter key. 

In another case, a user can think of a long word to switch state from B to C and navigate in the same directory level by thinking of a short word for selecting the right arrow and a long word for choosing the down arrow. Then the system goes to state A. It is possible that the classifier has made a mistake or that the user wants to go up the directory tree. Hence, in the state A, a user thinks of a long word to change state from A to D and thinks of a short word to revert to the previous action or a long word to go a level up in the tree. In this way, this design provides navigation among directories in the computer system and provides a simple way for computer interaction.

If a user is in state C and thinks either a short or long word to move in the same directory, then corresponding action (move right or bottom in the same hierarchy) is performed by the system. After performing the action, system transition occurs from state C to A without taking any input from the user, which is represented by the $\epsilon$ transition in the Figure \ref{stateDiagram2}. Similarly in state D, if the user thinks of a short or long word, then corresponding action (undo the previous action in the same level or move one level up in hierarchy) is performed and the system state is changed from D to A without taking any input from the user (denoted by $\epsilon$ transition).

\subsubsection{Design specifics}
Two designs presented here alternate between user input for 1 second and user rest state for 1 second. Here the maximum time is consumed in taking user input. Another 1 second is taken so that the user can decide to navigate within a directory. After waiting for the initial 1 second, the system pre-processes the signal, extracts useful features, classifies it to one of the categories, and takes appropriate action according to both: the classifier output and design implementation (design1 or design2). All the processing steps can be performed in milliseconds by the computer except for taking the user input.

\subsection{Implementation Details}
GUI implementation (design1) for displaying rectangles is performed using the Tkinter library in Python. Before starting the GUI, we train the classifier on the training data (as discussed later in \S \ref{sec:data_approach}). In this implementation, 60\% of the dataset was used for training the model parameters, and then the remaining 40\% of the data was used during the testing state. At the starting of each step, the system displays an option to select between crop and switch action and then alert the user to start thinking. The system then waits for 1 second to capture the EEG recording. In the partial-online analysis, a user clicks using the mouse in one part of the rectangle to select between a short and long word. After that, one trial of either short or long word from test data is selected at random. Selection from test data is based on the location of the click in the rectangle. If the location of the click is inside the top or left part of the rectangle, then a random test trial from the set of short words is selected. Otherwise, a random test trial from the set of long words is selected. Finally, the EEG signal is then pre-processed, transformed, and finally decoded by the classifier. All these steps are also repeated for taking inputs for the other states of FSM.

Figure \ref{crop} shows the display, rectangle division, and the target folder \textit{Downloads} (in blue) by two repeated crop actions. By cropping the rectangle, the user reaches the target location and selects the switch window option to double click on the desired folder. The window in which the rectangle is shown has been kept partially transparent so that the user can visualize the target folder's location and crop the rectangle accordingly.

\begin{figure*}[t]
    \centering
    \includegraphics[width=0.9\linewidth, height=7cm]{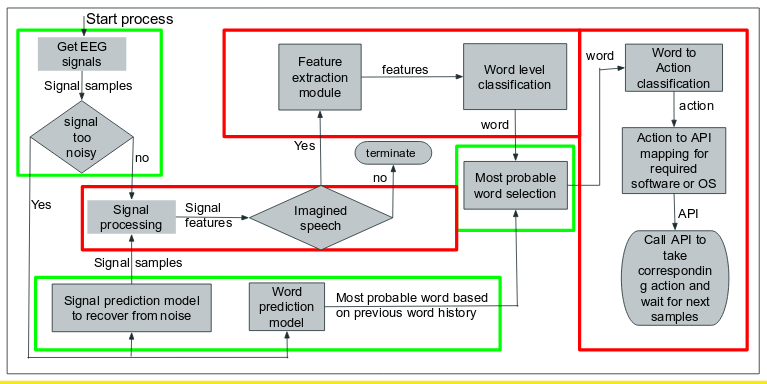}
    \caption{Pipeline for computer control using IS signals.}
    \label{flowGraph}
\end{figure*}

\subsubsection{GUI Design Considerations}
Many techniques can improve GUI performance as this design provides only a starting point for creating a real-time computer control system using imagined speech. 

\textbf{1)} Dividing the rectangle into multiple parts instead of two. A higher number of divisions imply that each step reduces the rectangle size by $k$ instead of 2. For example, if $k$ = 4 then, this new method is twice faster than the method with $k$ = 2. However, a higher value of $k$ requires a high accuracy of the multiclass classifier. 

\textbf{2)} When a classifier performs misclassification, then care has to be taken to circumvent this situation. For this situation to be rectified, the switch option is preferred. If a user detects that the last crop was incorrect, then in the next step, the user decides to switch to the previous rectangle. The switch action regenerates the previous rectangle, and the user can crop the rectangle in the next step. 

\textbf{3)} Prior to each crop with crop/switch state, the decision to crop or switch can be skipped for $k$ steps, and the value of $k$ can be decreased with each decision. 

\textbf{4)} This design considers only opening a folder by performing a double mouse click operation. Other options can also be provided for feature enhancement, such as a single mouse click, right-click of a mouse, and then creating a new window dictated from the size of the right-click menu. These features take the BCI system towards practical realization. 

\textbf{5)} This implementation is done in the Linux Operating System (OS), which means a few components are OS-dependent. Implementations can be made OS independent or developed for multiple operating systems.

\subsection{BCI Pipeline}
Based on the proposed approach, we can design a BCI system that identifies the rest state brain condition from the IS condition. If the brain signal corresponds to the IS condition, then it can decode the target word. Here, we have two modules performing two different categorization tasks. Similarly, a BCI system has other components related to artifact detection/removal and for OS interaction. By combining all the components, we propose a data flow framework of the IS-based BCI system (Figure \ref{flowGraph}). This framework is essential for the real-time functioning of the IS-based BCI system. The detail of each component is as follows.

After reading brain signals from the EEG device, it is necessary to identify whether the given signal is corrupted from the noise. If the signal is not corrupted, then the useful frequency components are extracted. After that, the filtered signal is examined to identify if the IS components are present within the signal. If the IS components are present, then the useful features are extracted, and the classification model is built to decode the imagined word. If the signal is noise corrupted, then a noise removal technique or signal reconstruction should be performed. A noisy signal also triggers the word prediction model. This model works based on the word's history. The classifier and word prediction model outputs are compared to identify the most probable word. This word is then mapped to the intended user action. Action is mapped to an application program interface (API). The API on execution changes the current state of the computer system. The modified system state again asks the user for some input and provides a new way of brain-computer interaction. As shown in Figure \ref{flowGraph}, we have implemented processes inside red boxes. Implementation of the processes inside green-colored boxes is left as a future work.

\section{Dataset Details and Decoding Imagined Speech Signal}\label{sec:data_approach}
In this section, we first provide details of the experiment and dataset.  Thereafter, we describe the proposed approach for decoding the EEG-based IS signals.

\subsection{Experiment and dataset details}
Nguyen et al. \cite{Nguyen2018IS} experimented with IS-related brain signals using an EEG device. Authors divided the experiment into four tasks namely short words $\{in,out,up\}$, vowels $\{a,i,u\}$, long words $\{independent, cooperate\}$ and short vs. long word $\{in, cooperate\}$.

\textit{Experiment}: In each experiment, subjects focus on a computer screen to receive the visual cue about the word to be imagined along with periodic beeps indicating the start of imagination. Each trial consists of 7 periods of $T$ seconds. Starting 4 periods consists of the visual cue with audio to imagine the word while the last three periods include only visual cue for imagined speech. The trial ended with 2 seconds of rest state condition without any beep sound or visual cue. Audio helped the subjects to estimate periodic intervals for imagining the pronunciation of vowels/words after the completion of 4 periods. For vowels and short words, $T$ is 1 second, and for long words and long vs. short word tasks, $T$ is 1.4 second. Each task has 100 trials for the target class corresponding to each subject.

\textit{Preprocessing}: EEG signals are captured using 64 electrodes and down-sampled to 256 Hz. Out of 64 channels, 60 channels were used for recording EEG signals of IS tasks. EEG signals are preprocessed by applying a 5th order Butterworth bandpass filter in the range 8-70 Hz, a notch filter at 60 Hz, and an algorithm of electrooculogram artifact removal \cite{he2004removal}. The dataset contains nine subjects for vowels IS task, six subjects for short words IS task, six subjects for long words IS task, and seven subjects for long vs. short word IS task. Each subject has 100 trials for every target class except for two subjects in the short vs. long word classification task having 80 trials each. We rejected data of one subject from short words IS task, three subjects from vowels IS task, one subject from long words IS task, and one subject from long vs. short word IS task due to a mismatch between the number of channels in the subject's data.

\textit{Dataset details}: Within each trial, subjects performed three repetitive thinking processes under the imagined speech condition. Hence, each trial gave rise to three different $[c,t]$ dimensional matrices with $c = 60,\;t = 256$ for vowels and short words and $ c = 60, \;t = 360$ for long words and short vs. long word tasks. For each subject, we have $[900,60,256]$ or $[600,60,360]$ dimensional matrix as input (except for two subjects in short vs. long words task where dimension is $[480,60,360]$) and $3$ or $2$-dimensional one hot vector as target labels depending on the $3$ vowels/short words category or $2$ long words/short vs. long word category.

\subsection{Approach Overview and Background}
We provide a brief description of our approach and an overview of the concepts used in the approach.

\subsubsection{Overview}
Our proposed approach for decoding the IS signal is summarized in the following steps. First, we create the covariance matrices from the raw EEG trials. Then, we project each of these covariance matrices to the tangent space (TS) to get a vector representation of the matrices. Third, we reduce the dimension of these vectors using PCA. Finally, features in the lower dimension are given as an input to the ensemble of NN classifiers, and the results of all classifiers are averaged to get the final prediction of the model. Based on the user's thoughts, the model makes a prediction, and after that, a corresponding action is performed on the computer screen for updating the user interface. Before a detailed explanation, we briefly go through the concepts used in our proposed approach.

\subsubsection{Covariance Matrix}\label{subsubsec:covmat}
Given an EEG trial $E \in \mathbb{R}^{n,m}$, covariance matrix $C \in \mathbb{R}^{n,n}$ is computed as $C = \frac{1}{m}E * E^T$, where $n$ is number of EEG channel and $m$ is number of samples, $T$ represents matrix transpose operation.


\subsubsection{Tangent Space}
In order to obtain a feature vector, a covariance matrix is transformed to the tangent space (ts) \cite{Barachant2012Riemann} as follows:
\begin{equation}\label{tsEq}
\begin{split}
P_i = C_m^{1/2}logm(C_m^{-1/2}C_iC_m^{-1/2})C_m^{1/2} \\
logm(M) = VD'V^{-1}, \; D'[i,i] = log(D[i,i])
\end{split}
\end{equation}
where $C_i$ is the covariance matrix, $C_m$ is the mean of the covariance matrices (denoted as reference point in Figure \ref{fig:ts+ann}),  $VDV^{-1}$ represents diagonalized form of the matrix $M$ and $P_i$ is the projected matrix.

\subsubsection{Principal Component Analysis}\label{subsubsec:pca}
In our work, we used PCA \cite{Jolliffe2014pca} for dimension reduction. The objective function of PCA is, $max_{u\in\mathbb{R}^n}\; u^TCu$ subject to $\|u\|_2^2 = 1$, where $C$ is the covariance matrix obtained and vector $u \in \mathbb{R}^n$.

\subsubsection{Artificial Neural Network}
We use Artificial Neural Network (ANN) \cite{Goodfellow2016DL}, \cite{ANN}, \cite{nnapprox} as a classification model. ANN linearly combines the input and then applies non-linearity (both steps applied in a layered fashion) to generate the desired output. Connectivity between two layers of ANN is defined as follows:
\begin{equation}\label{annEq}
a^l = g^l(W^l*a^{l-1})
\end{equation}
Here, vector $a^{l-1} \in \mathbb{R}^n$ represents an input obtained from layer $l-1$, $a^l \in \mathbb{R}^m$ represents an output at layer $l$, $W^l \in \mathbb{R}^{m,n}$ is the weight matrix between layer $l-1$ and $l$ and $g^l$ is the elementwise non-linear activation function at the layer $l$.

\subsubsection{Bootstrap Aggregation}
To increase the accuracy, we use the Bootstrap Aggregation (Bagging) \cite{Breiman1996bag} with ANN as a base classifier. This classification method creates several base classifiers and trains each on a subset of the original dataset. The result of the Bagging classifier is the average of base classifiers.

\subsection{Proposed Approach}

\begin{figure*}[t]
    \centering
    \includegraphics[width=0.9\linewidth]{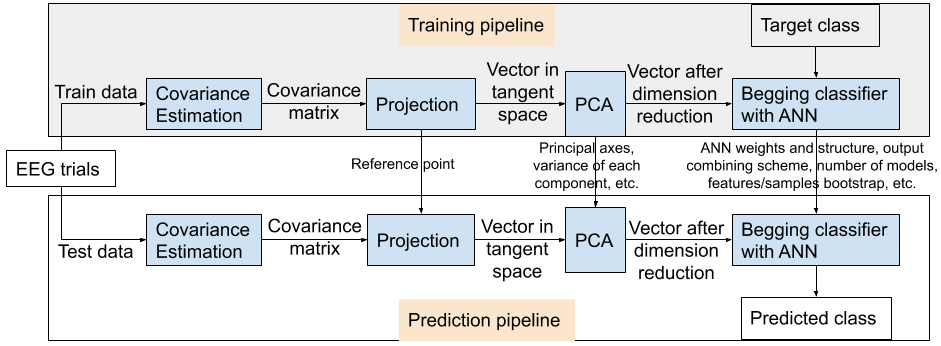}
    \caption{Proposed approach.}
    \label{fig:ts+ann}
\end{figure*}

In this section, we describe our proposed approach in detail (as shown in Figure \ref{fig:ts+ann}).

\subsubsection{Feature Extraction}
The following steps provide details of the proposed approach for feature extraction from the raw EEG signals.
\begin{itemize}
\item We store raw EEG trials in the format $[n,c,s]$ where $n$ is the number of trials, $c$ is the number of channels, and $s$ is the number of samples. Thereafter, we divide data into training and test set in the form $[n_{tr},c,s]$ and $[n_{te},c,s]$ where $n_{tr}$ and $n_{te}$ represents the number of trials in train and test set, respectively.

\item For each trial in the train and test set, covariance matrices are calculated (as described in section \ref{subsubsec:covmat}) and stored in the form $[n_{tr},c,c]$ and $[n_{te},c,c]$. After that, the train set is used to find the mean of covariance matrices, which is represented as $C_m$ in equation (\ref{tsEq}).

\item Each trial of the training and testing data is then projected to tangent space (as defined in equation \ref{tsEq}). After that projected matrices are converted to vector representation by concatenating rows of the matrix to form the matrices of dimension $[n_{tr},n_f]$ and $[n_{te},n_f]$ where $n_f$ denotes the number of features.

\item Feature dimension is then reduced by PCA (as defined in section \ref{subsubsec:pca}). The training data $[n_{tr},n_f]$ is used for learning the projection vectors ($u$). Thereafter, dimension of training and testing data is reduced using the learned vectors $u$ to form the matrices of dimension $[n_{tr},n_{rf}]$ and $[n_{te},n_{rf}]$, where $n_{rf}$ denotes the number of features obtained after the dimension reduction.
The number of features ($n_{rf}$) are selected from the set \{4, 8, 16, 32, 64\} using the cross-validation. 
\end{itemize}

The computation of obtaining covariance matrices from raw EEG signals and covariance matrix transformation to vectors was performed using the Pyriemann library \cite{pyriemann}. PCA implementation of sklearn \cite{sklearn} is used to project vectors into lower-dimensional space.

\subsubsection{Feature Classification}
To classify the features obtained after the dimension reduction, we utilize a bagging classifier with ANN as its base classifier. A bagging classifier contains $k$ base classifiers, where $k$ is a hyperparameter selected from the set \{2, 4, 8, 16, 32, 64\} using the cross-validation. The output of the bagging classifier is the average of $k$ ANN's.

The base classifier (ANN) takes $[n_b,n_{rf}]$ dimensional matrix as input and generates $[n_b,n_o]$ dimensional matrix as an output by applying equation (\ref{annEq}) in a layered fashion. In the matrix dimension, $n_b$ denotes the batch size, and $n_o$ denotes the number of output classes. The intermediate features of ANN are represented by $[n_{nb},n_f^l]$ where, $n_f^l$ denotes the number of features at layer $l$. In the implementation, we have used a single hidden layer, and the number of neurons in the hidden layer are selected from the set \{8, 16, 32, 64, 128, 256\} by performing the cross-validation. Non-linearity in the ANN is introduced by applying the ReLU activation function at the hidden layer.

At the output layer, the cross-entropy loss is computed through the ANN output matrix $O_{pred}$ of dimension $[n_b,n_o]$ and target output represented by one hot matrix $O_{true}$ of dimension $[n_b,n_o]$. Each row of the output matrix $O_{pred}$ generates target class probability and sums to 1. Each row of one hot matrix $O_{true}$ has all zeros but one for the class in which input belongs. For the regularization, the l2 penalty is applied with a regularization parameter of 0.0001. The gradients of the weights are calculated with respect to cross-entropy loss and computed using the back-propagation algorithm. ANN weights are updated using the gradient descent variant Adam optimizer \cite{Kingma2014adam}. The learning rate of the Adam optimizer is initialized with 0.001. At each iteration of the training, samples are randomly shuffled and divided into a mini-batch size of 200. The weights of ANN are initialized using the approach suggested by Glorot et al. \cite{glorot2010understanding}. The bagging classifier's training (with ANN as a base classifier) is performed using the sklearn library \cite{sklearn} of Python. To obtain each ANN weights' after the training, a class \textit{BaggingClassifier} in sklearn contains an attribute \textit{estimator} that provides the weights and hyper-parameters of each ANN.

\begin{figure*}[t]
    \centering
    \includegraphics[width=0.9\linewidth]{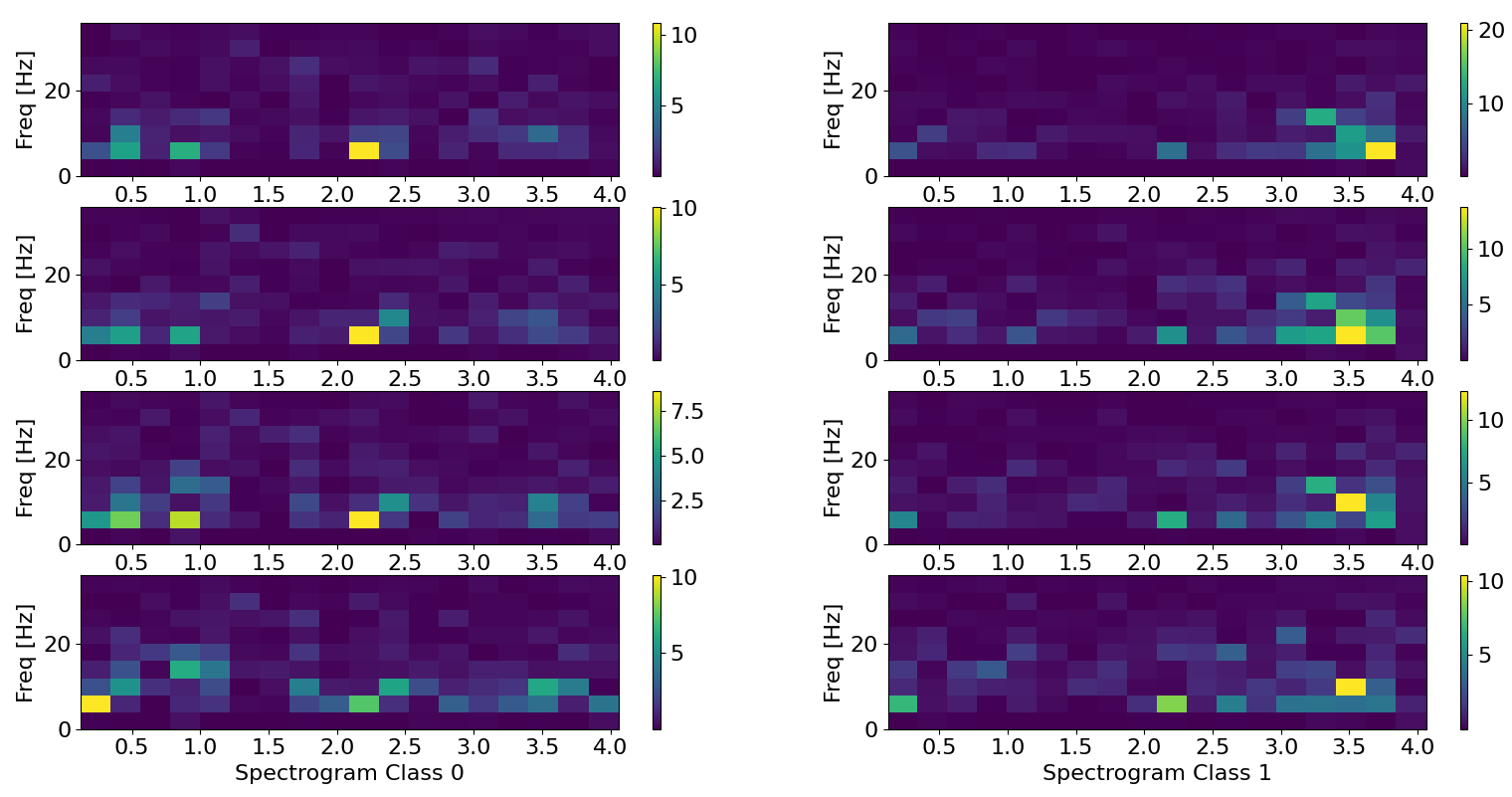}
    \caption{Spectrogram for two classes of the long words classification task.}
    \label{fig:spectrogram}
\end{figure*}

\section{Results}\label{sec:results}
This section shows the results of our proposed approach and then compares the results with various state-of-the-art approaches.

\subsection{Model Evaluation}

In this section, we first report results based on our proposed approach of using the covariance matrix, tangent space, PCA, and ANN with the begging classifier (termed as ts+ann in short); after that, compare with existing approaches for decoding the IS task.

\subsubsection{Performance metric}
We used classification accuracy $(CA)$ to check model performance. $CA$ measures the number of predicted outputs equal to actual outputs divided by the number of predictions. This quantity lies between 0 and 1. $(1-CA)$ denotes the model's misclassification rate. A train and test set is created by utilizing stratified 10-fold validation, which preserves the percentage of samples present in each class. Also, samples of each class are shuffled before dividing the data into batches for creating k-folds. 

\begin{figure*}[t]
    \centering
    \begin{subfigure}[t]{0.43\textwidth}
        \includegraphics[width=\textwidth]{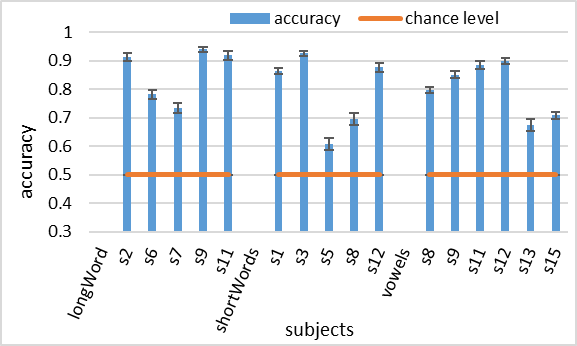}
        \centering
        \caption{}
        \label{restVsIS}
    \end{subfigure}%
    \begin{subfigure}[t]{0.57\textwidth}
        \includegraphics[width=\textwidth]{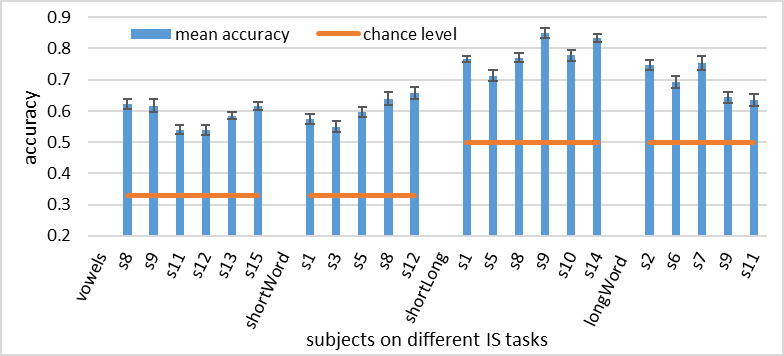}
        \centering
        \caption{}
        \label{fourTasksISperf}
    \end{subfigure}
    \caption{(a) Classification accuracy on long words imagined speech, short words, and vowels vs. rest state brain signals. (b) Classification accuracy of our proposed approach (ts+ann) on four different IS tasks. The participant’s id follows the name of each task. Error bars show the standard error of mean.}
    \label{}
\end{figure*}

\subsubsection{Feature analysis}
To analyze the IS-based EEG signal, we plot a spectrogram (in Figure \ref{fig:spectrogram}) of subject s11 corresponding to the two long-word classification task.
In Figure \ref{fig:spectrogram}, Class 0 and 1 corresponds to the two target classes of the long words classification task, and each row corresponds to a channel of the EEG signal.
The spectrogram shows that the IS-based EEG signals of two classes in the long-word classification task contain discriminative information, which a classifier can utilize to identify an imagined word.  

\subsubsection{IS vs. rest state}
We first show results for the classification of IS signals from the rest state signals. For comparison, we extracted IS signals of the 5\textsuperscript{th} period from the dataset and 2 seconds of rest state brain signals. We choose the 5\textsuperscript{th} period because it only contains visual cues for imagined speech condition, thereby avoiding any effects generated due to the audio. Using experimental results, we show that our proposed approach can separate IS signals from brain rest state signals with very high accuracy. We show these results on the classification task of long words, short words, and vowels in Figure \ref{restVsIS}. The high accuracy of many subjects on three different tasks shows that our proposed approach can successfully differentiate IS signals from rest state brain signals.

We performed a significance test of our proposed approach with chance level classification accuracy. We report p-values using 2 tailed t-test in Table \ref{pvalrestvsIS}. Small p-values show that results obtained using our proposed approach are significantly different from the chance level classification accuracy. 

\begin{table}[h]
\caption{P-Values for the proposed approach and chance level accuracy.}
\centering
\begin{tabular}{|c|c|} 
 \hline
 \rowcolor{lightgray}
 t-test & p-values \\
 \hline
 Vowels & 0.0005 \\
 \hline
 Short words & 0.0083 \\
 \hline
 Long words & 0.001 \\
 \hline
\end{tabular}
\label{pvalrestvsIS}
\end{table}

In Table \ref{meanstdRestVsIS}, we report the mean classification accuracy and standard deviation of all the subjects calculated for each classification task.
\begin{table}[h]
\caption{Classification accuracy across subjects on different IS classification tasks and rest state brain signals.}
\centering
\begin{tabular}{|c|c|c|} 
 \hline
 \rowcolor{lightgray}
 IS task vs. rest state & Mean accuracy & Standard deviation \\
 \hline
 Vowels & 0.8033 & 0.0858 \\
 \hline
 Short words & 0.794 & 0.1355 \\
 \hline
 Long words & 0.858 & 0.0927 \\
 \hline
\end{tabular}
\label{meanstdRestVsIS}
\end{table}
\begin{table}[h]
\caption{Mean, Standard Deviation (Std), Standard Error Of Mean (Sem), Maximum (Max) And Minimum (Min) Classification Accuracy for all Subjects on Different IS Tasks.}
\centering
\begin{tabular}{|m{1.1cm}|c|c|c|c|c|} 
 \hline
 \rowcolor{lightgray}
 Long words vs. rest & MEAN & STD & SEM & MAX & MIN \\
 \hline
 s2 & 0.9125 & 0.0406 & 0.0135 & 0.95 & 0.825 \\
 \hline
 s6 & 0.7825 & 0.0461 & 0.0153 & 0.875 & 0.7 \\ 
 \hline
 s7 & 0.735 & 0.0538 & 0.0179 & 0.825 & 0.675 \\
 \hline
 s9 & 0.94 & 0.0254 & 0.0084 & 0.975 & 0.875 \\
 \hline
 s11 & 0.92 & 0.0471 & 0.0157 & 0.975 & 0.8 \\
 \hline
 \rowcolor{lightgray}
 Short words vs. rest & MEAN & STD & SEM & MAX & MIN \\
 \hline
 s1 & 0.8633 & 0.0296 & 0.0098 & 0.9166 & 0.8166 \\
 \hline
 s3 & 0.9266 & 0.0249 & 0.0083 & 0.95 & 0.8666 \\
 \hline
 s5 & 0.6083 & 0.0597 & 0.0199 & 0.7166 & 0.55 \\
 \hline
 s8 & 0.695 & 0.0628 & 0.0209 & 0.7666 & 0.5333 \\
 \hline
 s12 & 0.8766 & 0.0454 & 0.0151 & 0.95 & 0.8166 \\
 \hline
 \rowcolor{lightgray}
 Vowels vs. rest & MEAN & STD & SEM & MAX & MIN \\
 \hline
 s8 & 0.7983 & 0.0292 & 0.0097 & 0.85 & 0.75 \\
 \hline
 s9 & 0.8516 & 0.039 & 0.013 & 0.9166 & 0.8 \\
 \hline
 s11 & 0.8866 & 0.042 & 0.014 & 0.9666 & 0.8 \\
 \hline
 s12 & 0.9 & 0.0324 & 0.0108 & 0.9666 & 0.8666 \\
 \hline
 s13 & 0.675 & 0.0606 & 0.0202 & 0.7666 & 0.55 \\
 \hline
 s15 & 0.7083 & 0.0389 & 0.0129 & 0.7666 & 0.6666 \\
 \hline
\end{tabular}
\label{meanstdsemRestVsIS}
\end{table}

A high accuracy of many subjects on three different classification tasks shows that our proposed approach can successfully differentiate IS signals from the rest state brain signals. Table \ref{meanstdsemRestVsIS} shows the mean classification accuracy, the standard deviation of the mean values, standard error of the mean, the maximum and minimum value of each subject in different IS tasks. Note that all subjects' minimum classification accuracy is well above the chance level for the long-word classification task compared to the other task of short words and vowels. Experimental results suggest that long words carry a lot more information than short words, which our proposed model uses to differentiate from the rest state brain signals.

\begin{figure*}[t]
    \centering
    \begin{subfigure}[t]{0.25\textwidth}
        \includegraphics[width=\textwidth,height=4cm]{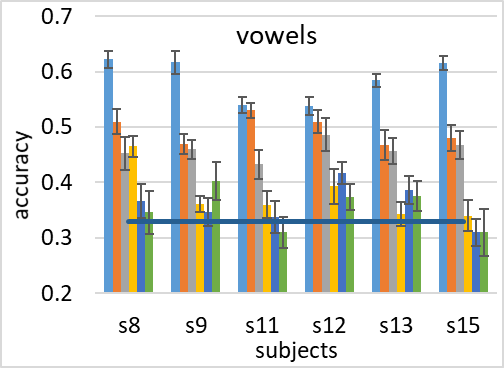}
        \centering
        \caption{}
        \label{vowels}
    \end{subfigure}
    \begin{subfigure}[t]{0.25\textwidth}
        \includegraphics[width=\textwidth,height=4cm]{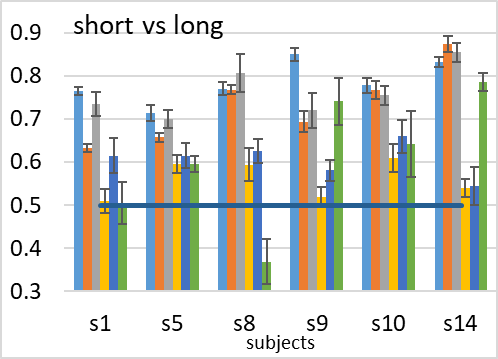}
        \centering
        \caption{}
        \label{shortLong}
    \end{subfigure}
    \begin{subfigure}[t]{0.28\textwidth}
        \includegraphics[width=\textwidth,height=4cm]{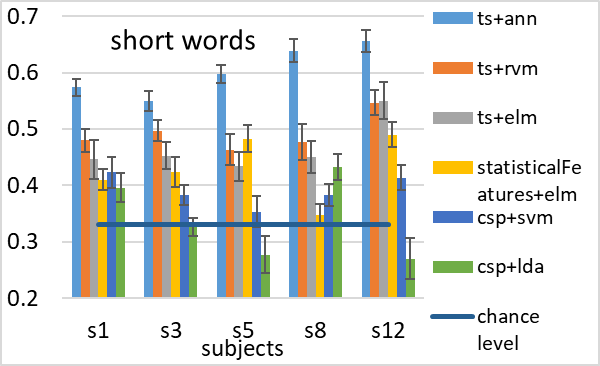}
        \centering
        \caption{}
        \label{short}
    \end{subfigure}
    \begin{subfigure}[t]{0.2\textwidth}
        \includegraphics[width=\textwidth,height=4cm]{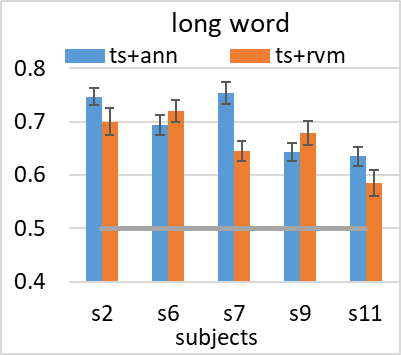}
        \centering
        \caption{}
        \label{long}
    \end{subfigure}
    \caption{Classification accuracy of different approaches on vowels, short vs long, short and long words classification tasks. Error bars show standard error of mean. The results for comparison are from the paper \cite{Nguyen2018IS}. Due to the unavailability of comparison results in long words IS task \cite{Nguyen2018IS}, we compare our approach only with ts+rvm.}
    \label{meanstd4IStasksFig}
\end{figure*}
\subsubsection{Four IS tasks}
Now we report results using our proposed approach on four IS tasks:  3 vowels, three short words, two long words, and one long vs. short word. 

Figure \ref{fourTasksISperf} shows the mean classification accuracy of the proposed approach on different subjects. We report a maximum mean classification accuracy of 0.85 for subject s9 on short vs. long word and a minimum of 0.5378 for subject s12 on vowels based IS classification task. Note that the classification accuracy is well above the chance level for each subject. High performance on short vs. long word classification task across all subjects states that the long word imagination leads to EEG patterns that are very different from the short word imagination. In Table \ref{meanstd4IStasksTab}, we report mean classification accuracy and standard deviation obtained on each IS task. Due to the complexity of short and long words, our approach obtains the highest mean classification accuracy in the short vs. long word IS task.

\begin{table}[h]
\caption{Mean Classification Accuracy and Standard Deviation on four IS tasks using our proposed approach of ts+ann.}
\centering
\begin{tabular}{|c|c|c|} 
 \hline
 \rowcolor{lightgray}
 IS task & Mean accuracy & Standard deviation \\
 \hline
 vowels & 0.586083 & 0.038881 \\
 \hline
 short words & 0.6035 & 0.044183 \\
 \hline
 short vs. long word & 0.785117 & 0.049689 \\
 \hline
 long words & 0.6943 & 0.055531 \\
 \hline
\end{tabular}
\label{meanstd4IStasksTab}
\end{table}

\begingroup
\setlength{\tabcolsep}{5pt} %
\begin{table*}[h!]
\caption{Detailed comparison with different approaches. Result are in the format: Mean Accuracy + Standard Deviation, Minimum Accuracy - Maximum Accuracy. The comparison results are from the paper \cite{Nguyen2018IS}, where results of the long words IS task are available only for ts+rvm approach.}
\centering
\begin{tabular}{|ccccccc|} 
 \hline
 \rowcolor{lightgray}
 \multicolumn{7}{|c|}{Vowels IS task}\\
 \hline
 Subjects & s8 & s9 & s11 &  s12 & s13 & s15 \\
 \hline
 \multirow{2}{8em}{csp+lda \cite{Blankertz2008}} & 34.6+11.8 & 40.3+10.4 & 31.0+8.5 & 37.3+7.1 & 37.5+8.1 &  31.0+12.7 \\
 & 16.6-60 & 13.3-46.6 & 20.0-46.6 & 30.0-53.3 & 26.6-53.3 & 10.0-50.0 \\
 \hline
 \multirow{2}{8em}{csp+svm \cite{Dasalla2009}} & 36.7+9.2 & 34.7+7.7 & 33.7+8.7 & 41.7+5.7 & 38.7+7.6 & 31.0+7.4 \\
 & 30.0-60 & 30.3-53.3 & 23.3-53.3 & 36.7-56.7 & 30.3-56.7 & 23.3-46.7 \\
 \hline
 \multirow{2}{8em}{statF+elm \cite{Min2016}} & 46.5+5.6 & 36.1+4.4 & 36.0+7.1 & 39.3+9.4 & 34.3+6.5 & 34.0+8.3  \\
 & 36.7-56.7 & 26.6-43.3 & 30.0-50.0 & 30.0-60.0 & 23.3-43.3 & 23.3-46.7 \\
 \hline
 \multirow{2}{8em}{ts+elm \cite{Nguyen2018IS}} & 45.3+8.9 & 46.0+5.1 & 43.3+7.9 & 48.6+8.9 & 45.7+7.2 & 46.7+7.5 \\
 & 30.0-56.7 & 36.7-53.3 & 33.3-53.3 & 36.7-60.0 & 36.7-63.3 & 36.7-60.0 \\
 \hline
 \multirow{2}{8em}{ts+rvm \cite{Nguyen2018IS}} & 51.0+6.7 & 47.0+5.5 & 53.0+4.0 & 51.0+6.3 & 46.7+8.2 & 48.0+7.2 \\
 & 43.3-63.3 & 36.7-53.3 & 46.7-60 & 43.3-63.3 & 33.3-60.0 & 33.3-56.7 \\
 \hline
 \centering
 \multirow{2}{8em}{ts+ann (proposed)} & \textbf{62.0}+4.68 & \textbf{61.66}+6.46 & \textbf{54.0}+4.16 & \textbf{53.78}+4.84 & \textbf{58.44}+3.52 & \textbf{61.55}+3.69 \\
 & 53.33-68.8 & 54.44-72.22 & 45.55-61.11 & 43.33-58.88 & 53.33-64.44 & 55.55-66.66 \\
 \hline
 \rowcolor{lightgray}
 \multicolumn{7}{|c|}{Short words IS task} \\
 \hline
 Subjects & s1 & s3 & s5 & s8 & s12 & \\
 \hline
 \multirow{2}{8em}{csp+lda \cite{Blankertz2008}} & 39.6+7.6 & 32.6+4.9 & 27.7+9.8 & 43.3+7.0 & 27+10.8 &\\
 & 26.6-53.3 & 26.6-43.3 & 20.0-50 & 36.6-53.3 & 13.3-43.3 &\\
 \hline
 \multirow{2}{8em}{csp+svm \cite{Dasalla2009}} & 42.3+8.2 & 38.3+5.3 & 35.3+8.3 & 38.3+6.1 & 41.33+6.7 &\\
 & 33.3-56.7 & 33.3-50.0 & 30.0-56.7 & 33.3-53.3 & 33.3-53.3 &\\
 \hline
 \multirow{2}{8em}{statF+elm \cite{Min2016}} & 41.0+5.5 & 42.3+8.0 & 48.3+7.2 & 34.7+5.9 & 49.0+6.7 &\\
 & 46.7-56.7 & 26.7-56.7 & 36.7-60.0 & 26.7-46.7 & 36.7-56.7 &\\
 \hline
 \multirow{2}{8em}{ts+elm \cite{Nguyen2018IS}} & 44.6+10.3 & 45.3+7.4 & 43.4+7.7 & 45.0+8.5 & 55.0+9.8 &\\
 & 33.3-60.0 & 33.3-56.7 & 30.0-56.7 & 30.0-56.7 & 40.0-70.0 &\\
 \hline
 \multirow{2}{8em}{ts+rvm \cite{Nguyen2018IS}} & 48.0+6.1 & 49.7+5.5 & 46.3+8.2 & 47.7+9.8 & 54.7+6.9 &\\
 & 40.0-56.7 & 40.3-56.7 & 36.7-66.7 & 36.7-66.7 & 43.3-66.7 &\\
 \hline
 \multirow{2}{8em}{ts+ann (proposed)} & \textbf{57.44}+4.55 & \textbf{55.0}+5.28 & \textbf{59.77}+4.91 & \textbf{63.88}+6.25 & \textbf{65.66}+5.99 &\\
 & 48.88-64.44 & 43.33-61.11 & 54.44-72.22 & 53.33-75.55 & 57.77-76.66 &\\
 \hline
 \rowcolor{lightgray}
 \multicolumn{7}{|c|}{Short vs long words IS task} \\
 \hline
 Subjects & s1 & s5 & s8 & s9 & s10 & s14 \\
 \hline
 \multirow{2}{8em}{csp+lda \cite{Blankertz2008}} & 50.5+14.8 & 59.5+5.7 & 36.9+15.9 & 74.1+16.6 & 64.3+23.0 & 78.5+6.3 \\
 & 30.0-72.5 & 52.5-70.0 & 21.9-71.9 & 31.3-87.5 & 20.0-80.0 & 70.0-90.0 \\
 \hline
 \multirow{2}{8em}{csp+svm \cite{Dasalla2009}} & 61.5+12.0 & 61.5+8.8 & 62.5+8.3 & 58.1+7.2 & 66.0+11.5 & 54.5+13.2 \\
 & 50.0-85.0 & 50.0-80.0 & 50.0-81.3 & 50.0-75.0 & 50.0-85.0 & 45.0-90.0 \\
 \hline
 \multirow{2}{8em}{statF+elm \cite{Min2016}} & 51.0+8.4 & 59.5+6.4 & 59.4+11.5 & 51.9+6.6 & 61.0+9.7 & 54.0+6.1 \\
 & 40.0-65.0 & 50.0-70.0 & 43.8-81.3 & 43.8-68.8 & 45.0-75.0 & 50.0-70.0 \\
 \hline
 \multirow{2}{8em}{ts+elm \cite{Nguyen2018IS}} & 73.5+8.2 & 70.0+6.2 & \textbf{80.6}+13.2 & 72.5+12.2 & 75.5+6.8 & 85.5+6.8 \\
 & 60.0-85.0 & 60.0-80.0 & 62.5-93.8 & 43.7-87.5 & 65.0-85.0 & 75.0-95.0 \\
 \hline
 \multirow{2}{8em}{ts+rvm \cite{Nguyen2018IS}} & 63.3+2.9 & 65.8+3.1 & 76.9+3.0 & 69.4+7.5 & 76.8+6.2 & \textbf{87.5}+5.5 \\
 & 60.0-70.0 & 62.5-70.0 & 71.8-81.3 & 59.4-81.3 & 67.5-85.0 & 75.0-92.5 \\
 \hline
 \multirow{2}{8em}{ts+ann (proposed)} & \textbf{76.5}+2.83 & \textbf{71.33}+5.66 & 77.08+4.26 & \textbf{85.0}+4.73 & \textbf{77.83}+5.16 & 83.33+3.57 \\
 & 73.33-81.66 & 60.0-78.33 & 68.75-85.41 & 77.08-91.66 & 68.33-85.0 & 78.33-90 \\
 \hline
 \rowcolor{lightgray}
 \multicolumn{7}{|c|}{Long words IS task} \\
 \hline
 Subjects & s2 & s6 & s7 & s9 & s11 & \\
 \hline
 \multirow{2}{8em}{ts+rvm \cite{Nguyen2018IS}} & 70.0+7.8 & \textbf{72.0}+0.6 & 64.5+5.5 & \textbf{67.8}+6.8 & 58.5+7.4 &\\
 & 55.0-80.0 & 65.0-85.0 & 59.0-75.0 & 55.0-80.0 & 50.0-77.5 &\\
 \hline
 \multirow{2}{8em}{ts+ann (proposed)} & \textbf{74.66}+4.76 & 69.33+5.53 & \textbf{75.33}+6.35 & 64.33+5.12 & \textbf{63.5}+5.39 & \\
 & 66.66-81.66 & 60.0-76.66 & 65.0-86.66 & 56.66-71.66 & 53.33-73.33 & \\
 \hline
\end{tabular}
\label{IStasksCompTab}
\end{table*}
\endgroup

\begingroup
\setlength{\tabcolsep}{1pt} %
\begin{table*}[t!]
\caption{P-Values Obtained After Two Tailed Paired T-Test. The comparison results are derived from the paper \cite{Nguyen2018IS}, where results of the long words IS task are available only for ts+rvm approach.}
\centering
\begin{tabular}{|c|c|c|c|c|c|c|} 
 \hline
 \rowcolor{lightgray}
 IS task & ts+ann, chance level & ts+ann, ts+rvm & ts+ann, ts+elm & ts+ann, statF+elm & ts+ann, csp+svm & ts+ann, csp+lda \\
  \hline
 vowels & 0.000016673 & 0.0117196 & 0.000780616 & 0.000246665 & 0.00038811 & 0.0000936821 \\
 \hline
 short words & 0.000157915 & 0.00385681 & 0.001383572 & 0.005317578 & 0.000641303 & 0.002537817 \\
 \hline
 short vs. long & 0.0000327993 & 0.159018237 & 0.372650079 & 0.001139683 & 0.002841504 & 0.019606281 \\
 \hline
 long words & 0.001440737 & 0.34204982 & - & - & - & - \\
 \hline
 \end{tabular}
\label{pval4IStasksTab}
\end{table*}
\endgroup

\subsubsection{Comparison}
\textit{Baseline}: We compare our approach with two baselines (CNN and RNN) on the three short words and two long words classification tasks. To model spectral and spatial features of the EEG signals, CNN contains five convolutions, three pooling, and two dropout layers. To reduce the number of training parameters, the output of the 5\textsuperscript{th} convolution layer is directly connected to the output layer of CNN. Nonlinearity in the network is introduced by applying the ReLU activation function between convolution and pooling layers. A dropout layer is added to reduce the network's overfitting and is applied after the pooling layers. The kernel size at each convolution layer and the dropout rate is a hyperparameter, which are tuned using cross-validation. 

To model the temporal nature of EEG signals, a variant of RNN named GRU \cite{gru} is used as a baseline. At each time step of the GRU, a sample of the EEG trial is provided as an input to the model. Similarly, the output is collected from the GRU at each time step. The output of all time steps is averaged to generate a single output of GRU.
Both CNN and RNN are trained using the minibatch size of 32 and using the Adam optimizer. The weights of CNN and RNN are initialized using the approach of Glorot et al. \cite{glorot2010understanding}.

\begin{figure}[t]
    \centering
    \begin{subfigure}[t]{0.5\linewidth}
        \includegraphics[width=\linewidth,height=4cm]{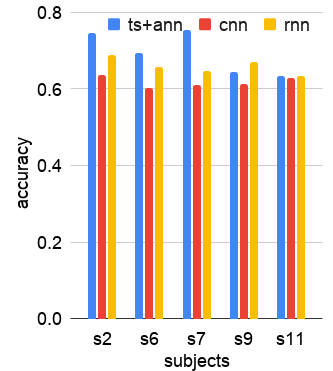}
        \centering
        \caption{}
        \label{BaselineLongWords}
    \end{subfigure}%
    \begin{subfigure}[t]{0.45\linewidth}
        \includegraphics[width=\linewidth,height=4cm]{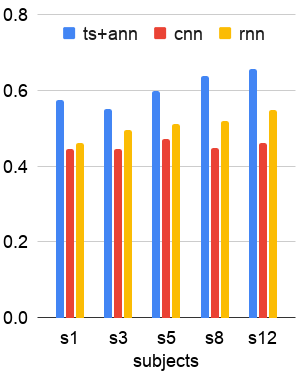}
        \centering
        \caption{}
        \label{BaselineShortWords}
    \end{subfigure}
    \caption{Classification accuracy of our proposed approach and two baseline approaches (CNN, RNN) on classification tasks: (a) 2 long words, (b) 3 short words.}
    \label{baselineCompare}
\end{figure}

Figure \ref{baselineCompare} compares the proposed approach with two deep learning baselines CNN and RNN. Figure \ref{baselineCompare} shows that the proposed approach of ts+ann outperforms both baselines on the short words classification task and performs better on three subjects (out of 5) on two long words classification task. These results confirm the generalization capability of the proposed approach of ts+ann over two baselines of CNN and RNN.

\textit{Existing approaches}: Now we compare our proposed approach of using ts+ann with existing approaches on different IS tasks  (Figure \ref{meanstd4IStasksFig}). We compare our approach of using ts+ann with: (a) ts as features with rvm as a classifier approach suggested by Nguyen et al. \cite{Nguyen2018IS}; (b) ts as features, and elm as a classifier approach also suggested by Nguyen et al. \cite{Nguyen2018IS}; (c) use of statistical features with elm as a classifier suggested by min et al. \cite{Min2016}; (d) CSP based transformed signal with SVM as a classifier approach of Dasalla et al. \cite{Dasalla2009}; and (e) variance of CSP transformed signal with LDA as a classifier suggested by Tomioka et al. \cite{Blankertz2008}.
The results for comparison are from the paper \cite{Nguyen2018IS}, which are available for three different IS tasks: vowels, short words, and short vs. long word.
Due to the unavailability of comparison results for the long-word classification task in \cite{Nguyen2018IS}, we compare our proposed approach only with the ts+rvm approach \cite{Nguyen2018IS}.

\textit{Detailed comparison}: As reported in Figure \ref{meanstd4IStasksFig}, our approach outperforms existing approaches in vowels, and short-words IS tasks and performs equivalent to the ts+elm, ts+rvm approach on decoding short vs. long word and two long words IS tasks, respectively. We report the highest classification accuracy of 0.85 on short vs. long word IS task for subject s9 and minimum classification accuracy of 0.5378 for subject s12 on vowels IS task. Results obtained using our approach of ts+ann are well above chance level for all subjects on all four IS tasks. Chance level accuracy is 0.33 for vowels, and short words and 0.5 for short vs. long word and long-word IS tasks. Due to the dimension reduction with PCA and the generalization capability of ANN, our proposed approach outperforms other approaches with a significant margin on vowels and short words IS tasks. For the short vs. long words classification task, our approach ts+ann outperforms the approach proposed by Nguyen et al. \cite{Nguyen2018IS} on four subjects (out of 6) with a significant margin.
Similarly, on the long words classification task, our approach ts+ann outperforms the approach proposed by Nguyen et al. on three subjects (out of 5). In our approach, the Bagging classifier helps in reducing the variance in predicting the output. Therefore, results are more stable in terms of SEM in comparison to the other approaches.

We compare the mean classification accuracy, standard deviation, maximum, and minimum accuracy obtained using our approach with existing approaches in Table \ref{IStasksCompTab}. For the long words IS task, we compared our approach only with the approach proposed by Nguyen et al. \cite{Nguyen2018IS} due to the lack of comparison results in \cite{Nguyen2018IS}.
As we observe from Table \ref{IStasksCompTab}, our proposed approach is able to outperform other approaches in short words and vowel classification tasks and performs equivalent with the approach suggested by Nguyen et al. \cite{Nguyen2018IS} on short vs. long words and long words classification task. Our proposed approach of using ts+ann also has less deviation in comparison to the other approaches. This is achieved by using an ensemble of ANN classifiers and averaging the results.

\textit{T-test}: Now we perform significance testing of our proposed approach with chance level accuracy and other approaches. Table \ref{pval4IStasksTab} shows the p-values after performing the two-tailed pairwise t-test. Results in Table \ref{pval4IStasksTab} show a very low p-value when comparing our proposed approach ts+ann with chance level accuracy. Hence, our approach performs well above the chance level on all four IS tasks.
In comparison to the approach proposed by Nguyen et al. \cite{Nguyen2018IS}, results obtained using our approach are significantly different for vowels and short words IS tasks. This is verified by the low p-values of 0.01171 and 0.00385 for a 0.05 significance level. In contrast, for short vs. long word and long words IS tasks, p-values 0.15901 and 0.34204 show the equivalence of results between our proposed approach (ts+ann) and Nguyen et al. \cite{Nguyen2018IS} (ts+rvm). Similar behavior is also observed for the ts+elm approach, which is suggested by Nguyen et al. \cite{Nguyen2018IS}.

For all other approaches, we see that p-values are far below the significance level. Hence, it shows that results obtained with our approach (ts+ann) are significantly different from approaches suggested by min et al. \cite{Min2016} (statistical features with elm as a classifier), Dasalla et al. \cite{Dasalla2009} (CSP based transformed signal with SVM as a classifier) and Tomioka et al. \cite{Blankertz2008} (variance of CSP transformed signal with LDA as a classifier) on vowels, short words and short vs. long word classification tasks.

\textit{Averaged performance per approach}: From Table \ref{IStasksCompTab}, we observe that the performance of each approach varies significantly across subjects. To compare different approaches, we require a result from each of the considered approaches. To this end, we average the performance of each approach across all subjects. This provides one performance measure for each classification task. Table \ref{meanstd4IStasksSubAvgTab} summarizes these results.
\begin{table}[t]
\caption{Mean classification accuracy (Mean) and standard deviation (Std) computed across all subjects for each IS task. The comparison results are derived from the paper \cite{Nguyen2018IS}, where results of the long words IS task are available only for ts+rvm approach.}
\centering
\begin{tabular}{|m{1.5cm}|m{0.7cm}|m{1cm}|m{1cm}|m{1cm}|m{1cm}|}
 \hline
 \rowcolor{lightgray}
 \multicolumn{2}{|c|}{Task} & Short words & Vowels & Short vs. Long & Long words \\
 \hline
 \multirow{2}{5em}{csp+lda} & Mean & 34.04 & 35 & 64.83 & \multirow{2}{3em}{-} \\
 & Std & 7.21 & 3.91 & 10.12 & \\
 \hline
 \multirow{2}{5em}{csp+svm} & Mean & 39.1 & 35.5 & 61 & \multirow{2}{5em}{-} \\
 & Std & 2.77 & 3.8 & 4.64 & \\
 \hline
 \multirow{2}{5em}{statF+elm} & Mean & 43.06 & 37.5 & 56.16 & \multirow{2}{5em}{-} \\
 & Std & 5.86 & 4.71 & 4.75 & \\
 \hline
 \multirow{2}{5em}{ts+elm} & Mean & 46.66 & 45.5 & 75 & \multirow{2}{5em}{-} \\
 & Std & 4.71 & 1.81 & 5.25 & \\
 \hline
 \multirow{2}{5em}{ts+rvm} & Mean & 50.1 & 49.0 & 73.3 & 66.2 \\
 & Std & 3.5 & 2.4 & 8.86 & 4.8 \\
 \hline
 \multirow{2}{5em}{ts+ann} & Mean & \textbf{60.35} & \textbf{58.6} & \textbf{78.5} & \textbf{69.43} \\
 & Std & 4.41 & 3.88 & 4.96 & 5.55 \\
 \hline
 \end{tabular}
\label{meanstd4IStasksSubAvgTab}
\end{table}
Table \ref{meanstd4IStasksSubAvgTab} shows that our proposed approach gives the highest accuracy across all the IS classification tasks. By examining the standard deviation of our approach, it is clear that the ANN model does not show much variability across subjects. Other approaches show low mean accuracy with low variance or high accuracy with high variance across all the subjects. Hence, existing approaches are either unable to extract useful discriminative information, thereby resulting in low accuracy and low deviation, or these approaches can decode IS signals of some subjects and, therefore, achieve high accuracy, however, with high variance. The approach with high mean accuracy and low variance (when calculated across all subjects) is desired. 

\textit{Kappa score}: In Table \ref{kappaIStasksSubAvgTab}, we show the kappa score for each classification task. To evaluate the kappa score, all subjects' accuracy is averaged and compared against chance level accuracy for each task. Kappa score is calculated as (accuracy with an approach - chance level) / (1 - chance level). If the accuracy of an approach is close to chance level, then the kappa score will be close to 0. Similarly, if an approach achieves an accuracy of 1, then the kappa score is 1.
In contrast, if the accuracy is below the chance level, then the kappa score is negative. From Table \ref{kappaIStasksSubAvgTab}, it is evident that the kappa score of the proposed approach is positive and higher than the existing approaches. Therefore, the relative accuracy of the proposed approach (ts+ann) w.r.t. chance level is much higher than the existing approaches. 

\begin{table}[t]
\caption{Kappa score computed across all subjects for each IS task. The comparison results are derived from the paper \cite{Nguyen2018IS}, where results of the long words IS task are available only for ts+rvm approach.}
\centering
\begin{tabular}{|m{1.5cm}|m{1cm}|m{1cm}|m{1.3cm}|m{1cm}|}
 \hline
 \rowcolor{lightgray}
 Task & Short words & Vowels & Short vs. Long & Long words \\
 \hline
 csp+lda & 0.015 & 0.02 & 0.29 & - \\
 \hline
 csp+svm & 0.09 & 0.03 & 0.22 & - \\
 \hline
 statF+elm & 0.15 & 0.06 & 0.12 & - \\
 \hline
 ts+elm & 0.203 & 0.18 & 0.5 & - \\
 \hline
 ts+rvm & 0.25 & 0.23 & 0.46 & 0.32 \\
 \hline
 ts+ann & 0.408 & 0.382 & 0.57 & 0.388 \\
 \hline
 \end{tabular}
\label{kappaIStasksSubAvgTab}
\end{table}

\textit{Conclusion}: The results show the generalization capability of ANN models over other models when given the same input data. Note that the classification accuracy of words in the long words IS task and short vs. long words IS task is much higher than vowels and short-word IS tasks. The high classification accuracy for long words and short vs. long words IS tasks suggests that the proper choice of words based on word length and complexity provides useful discriminative information and improves the models' generalization power.

\begin{figure*}[t]
    \centering
    \begin{subfigure}[t]{0.3\textwidth}
        \includegraphics[width=\textwidth,height=4cm]{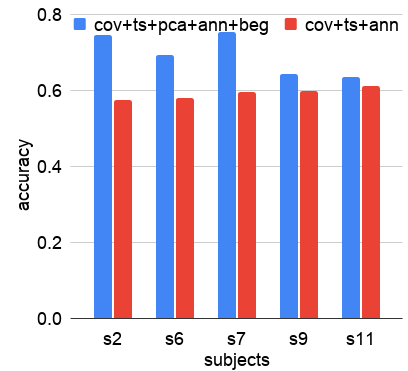}
        \centering
        \caption{2 long words}
        \label{ablationLongWords}
    \end{subfigure}%
    \begin{subfigure}[t]{0.3\textwidth}
        \includegraphics[width=\textwidth,height=4cm]{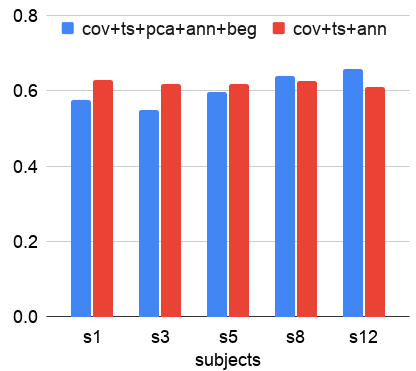}
        \centering
        \caption{3 short words}
        \label{ablationShortWords}
    \end{subfigure}%
    \begin{subfigure}[t]{0.3\textwidth}
        \includegraphics[width=\textwidth,height=4cm]{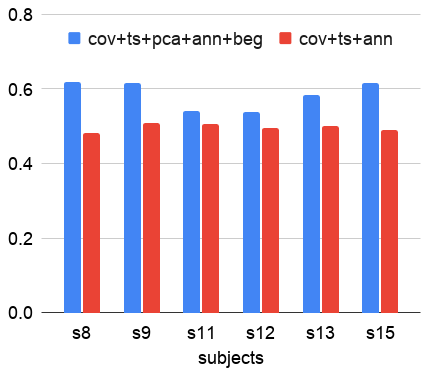}
        \centering
        \caption{3 vowels}
        \label{ablationVowels}
    \end{subfigure}
    \caption{Ablation results with and without pca and bagging classifier. Classification accuracy of our proposed approach cov+ts+pca+ann+bag and a second approach of cov+ts+ann on different IS classification tasks.}
    \label{ablationCompare}
\end{figure*}

\subsubsection{Ablation results}
We perform the ablation experiment to demonstrate the efficiency of both PCA and Bagging classifier in the proposed approach. Figure \ref{ablationCompare} compares the proposed model's accuracy (covariance matrix, tangent space, PCA, and Bagging with ANN as a base classifier) and a model using the covariance matrix, tangent space, and ANN as a classifier. Figure \ref{ablationCompare}, shows that the PCA and Bagging classifier indeed improves the model's accuracy in the two long words and three vowels classification task. The high accuracy with the proposed approach shows that the features obtained after PCA contain discriminative information and reduce the number of training parameters of the ANN, thus avoiding overfitting of the model.  In the three short-word classification task, the performance drop is observed for three out of 5 subjects. A reason for the performance drop is the loss of discriminative information after applying the dimension reduction on the features obtained in the tangent space.  Based on the results shown in Figure \ref{ablationCompare}, it is advised to decode IS signals belonging to the long words and vowels category with the proposed approach. For the short words category, EEG signal classification for a subject should be performed with and without PCA, and a model with the highest validation accuracy should be selected. 

\subsubsection{Design evaluation}
To evaluate an FSM design for the IS based BCI system, we compute average information (denoted as $I$) across different trials of a subject as follows:
\begin{equation}\label{infoEq}
I = log_2|C| + alog_2a + m log_2\frac{m}{|C|-1}
\end{equation}

Where, $|C|$ is the number of classes in the target class set $C$, $a$ is the classification accuracy and $m$ is the misclassification rate of the classifier computed as $1-a$. 
Now information transfer rate (ITR) can be obtained as, 
\begin{equation}\label{itrEq}
ITR =I/T
\end{equation}

Where, $I$ is average information in bits per trial and $T$ is the total time of each trial.
For our analysis, we have $|C|=2$ since $C=\{0,1\}$,  $a=0.95$, $m=0.05$ and $T=2$ seconds. 
Hence, $ITR$ in our case is 0.35 bits per second or 21 bits per minute. As shown later in the results section, $a=0.95$ is the mean classification accuracy of subject s9 rounded to the nearest decimal (0.94 is upto 2 decimal places) for the classification task of rest state brain signals vs. long word IS signals (\S \ref{sec:results}).
\section{Discussion, Related Work and Conclusion}
This section discusses a few points related to performance on different datasets and design aspects of an IS based BCI system. After that, we provide the concluding remarks. 

\subsection{Discussion}
In this paper, we have shown that our proposed approach can decode IS-based brain signals such as long words and short vs. long word with high accuracy. Also, our approach generalizes to vowels and short words IS tasks by improving the results of the state-of-the-art approaches. We observe that an appropriate feature and a classification model can improve the word decoding capability. Because words or vowels are different in their speech signal representation, the process generating these sounds inside the brain must generate different activation patterns. These activation patterns lead to the discriminative EEG signals. Since long words are more complicated than vowels and short words in terms of imagined pronunciation. Hence, this additive complexity in silently speaking the long words provides more discriminative information and improves the classification results for short vs. long word and long words decoding tasks.

The behavior of sophisticated features representation is also supported by the results obtained for the IS signal vs. rest state brain signal. We observe high classification accuracy for long words vs. rest state in comparison to short words/vowels vs. rest state brain signals. We observe a difference of around 5\% in classification accuracy when decoding long words IS signals from rest state in comparison to vowels/short words IS signals from rest state brain signals. We believe that model performance will increase if the time difference of capturing IS signal and rest state signal increases. In the experiment by \cite{Nguyen2018IS}, the rest state condition was immediately followed by the IS condition. So there is some chance that subjects are still in an imaginary state, and thus feature representation is the same.

\textit{Extension of the interface design}: The proposed work also presents two new interface designs for computer control with the following benefits and differences from existing designs: \textbf{1)} This interface design is generic. Though we illustrate for IS-based EEG tasks, the design can be expanded for use in other BCI paradigms such as motor imaginary. However, the interface currently being used in P300 speller or motor imaginary to control the mouse cannot be easily used in IS tasks. \textbf{2)} The interface in this paper is shown for the binary classification tasks, but it can easily be extended to a multiclass setup for providing faster navigation or multiple features to the user. \textbf{3)} Design 2 provides an easy way of navigation within a file structure. Since design 1 is generic, it can be used in various computer applications such as folder navigation, browser-control, or navigating through documents (reading purpose). \textbf{4)} Designs 1 and 2 can be easily extended for providing more functionality to users, such as providing right-click, double-click, and single-click of the mouse. \textbf{5)} One crucial difference between existing designs and proposed design is that designs presented in this work are reactive instead of existing proactive designs.

\textit{Reactive vs. proactive design}: In the case of reactive design, we wait for the user signal to modify the current system. In contrast, the proactive design keeps the system active by automatically moving the cursor over available computer screen options. The user requires to provide input when the target location is reached. An example of proactive design is the horizontal line movement from top to bottom and a vertical line movement from left to right on the computer screen. These lines create various intersection points on the computer screen. When the intersection of lines is at the target location, the user provides an input, and the system state changes accordingly. After that, the whole process repeats. The second proactive design is a circular rotation of a line segment starting from a computer screen center up to the end of a computer screen. When the line intersects with a target location, such as some folder, the user provides input, and the line rotation stops (say at an angle of degree $\theta$ from positive-x/horizontal direction). Then line segments of different lengths are displayed from small lengths up to the maximum screen resolution size along the direction $\theta$. A user provides a second input whenever a new line segment reaches the target location. The above are few examples of proactive interface design to operate the computer system. 

\textit{Design simulation}: Note that the designs presented here are tested in a partial online setting. In a partial online setting, rather than taking input from an EEG device, the input was taken from a user mouse click. Based on the click's location, corresponding IS based EEG trial from the test set was picked, processed, classified, and the system state was changed. The new system state was shown to the user, and then the user again provides input to reach a target location. This provides a closed-loop of the BCI system for human-computer interaction. 

\subsection{Related work}
Now we discuss existing approaches for decoding IS signals. These approaches are state of the art and widely used in decoding brain signals.

Nguyen et al. \cite{Nguyen2018IS} used 64 channel EEG to capture three vowels, two long words, and three short words across different subjects. They used features from Riemann manifold (Tangent Space (TS)) \cite{Barachant2012Riemann} as an input to the Relevance Vector Machine (RVM) \cite{Psorakis2010RVM}. Their results show the mean classification accuracy of 49\% for vowels and short words, 66\% for long words across, and 73\% for long vs. short word classification tasks for different subjects. We significantly improve these results for vowels and short-word classification tasks and obtain equivalent results for long words and short vs. long word classification tasks. This improvement was achieved by reducing the dimensions of the transformed covariance matrix using the PCA and a more powerful NN classifier combined with bagging as an ensemble classifier. The authors also suggest using Extreme Learning Machine (ELM) as a classifier, but they obtained very similar results by using RVM as a classifier.

Tomioka et al. \cite{Blankertz2008} applied Common Spatial Pattern (CSP) to data, calculated log-variance of each transformed channel to create input features, and used linear discriminant analysis (LDA) as a classifier. Here, the LDA classifier poses the main limitation, which works well if the features of each class are generated using the normal distribution. Our approach removes this limitation by using a robust ANN classifier for modeling complex distributions.

Dasalla et al. \cite{Dasalla2009} used CSP-based transformation and support vector machine (SVM) as a classifier. They used four CSP channels for transforming raw EEG signals using the training data and then transformed both training and test data using learned parameters. Signals obtained after transformation were stacked together to form a vector and finally given as input features to the SVM classifier. The authors used CSP, which is known to work best for the case of motor imaginary signals. In contrast, we used the covariance matrix as input features to our model, capturing the dependence between different channels and retaining some information caused by imagined speech production.

Min et al. \cite{Min2016} used statistical features such as mean, variance, standard deviation, and skewness, and ELM as a classifier. To extract these features, they divided the signals into overlapping windows and calculated these features over each channel of each window to form a feature vector, used a sparse-regression-based feature selection scheme to reduce the dimension of the features, and used ELM as a classifier. In our proposed approach, we used PCA for feature dimension reduction as this method is much faster than a sparse regression-based feature dimension approach. Another difference lies in the use of the classification model. Our feed-forward neural network model is trained using gradient descent, and gradients are computed using a backpropagation algorithm (rather than random initialization of weights in the first layer of ELM). Gradient descent makes our model more robust and shows good generalization on test data.

In other contexts such as motor imagery classification of EEG signals, Amin et al. \cite{amin} proposed a deep learning model based on convolution neural networks.
Acharya et al. \cite{acharya} provided a review of the EEG signals based on the focal and non-focal category to detect brain areas affected by seizures.
Weijie et al. \cite{weijie} proposed various signal processing techniques for analyzing the brain signals. 
Rajdeep et al. \cite{rajdeep} proposed a feature selection approach (based on the machine learning technique) for the motor imagery signal.
Hyun et al. \cite{Hyun} proposed an approach based on genetic algorithms and neural networks to recognize Alzheimer's disease based on the EEG signals.

\subsection{Conclusion}

Our proposed work shows that it is possible to design an imagined speech signal-based brain-computer interface for human-computer interaction with machine learning techniques. We have presented an approach using the covariance matrix as input feature and ANN as classification model for decoding IS signal. This approach outperformed existing methods when applied on decoding short words and vowels across all the subjects. Similarly, for long words and long vs. short words classification tasks, our approach outperformed existing approaches on the majority of subjects. We show that IS signals can be differentiated from other brain signals, and the length of words is a useful criterion in discriminating words. In the future, we will work on improving model performance, developing new ways of computer interaction, and IS signal prediction models to recover from high noise scenarios.

\bibliography{reference}

\end{document}